\documentclass[12pt]{article}

\usepackage{epsf}
\usepackage[dvips]{graphicx}
\usepackage[dvips]{graphics}

\let\ensm=\ensuremath
\def\Pp{{\rm p}}

\def\PK{{\rm K}}

\def\Pu{{\rm u}}
\def\Pd{{\rm d}}
\def\Ps{{\rm s}}

\def\pip               {\ensm{\pi^+}}
\def\pim               {\ensm{\pi^-}}
\def\pipm              {\ensm{\pi^\pm}}

\def\gevp             {\ensm{\,{\rm GeV}/c}}

\begin{document}

\begin{center}
{\bfseries SINGLE SPIN ASYMMETRY IN HIGH P$_{T}$ CHARGED HADRON PRODUCTION OFF
 NUCLEI AT 40 GEV}

\vskip 5mm

 V.V.~Abramov$^{1 \dag}$, P.I.~Goncharov$^{1}$, A.Yu.~Kalinin$^{1}$,
 A.V.~Khmelnikov$^{1}$, A.V.~Korablev$^{1}$,  Yu.P.~Korneev$^{1}$,
 A.V.~Kostritsky$^{1}$, A.N.~Krinitsyn$^{1}$, V.I.~Kryshkin$^{1}$,
 A.A.~Markov$^{1}$,   
V.V.~Talov$^{1}$,  L.K.~Turchanovich$^{1}$ and
  A.A.~Volkov$^{1}$

\vskip 5mm

{\small
(1) {\it
Institute for High Energy Physics, Protvino, Moscow region, Russia
}
\\
$\dag$ {\it
E-mail: Victor.Abramov@ihep.ru
}}
\end{center}

\vskip 5mm

\begin{center}
\begin{minipage}{150mm}
\centerline{\bf Abstract}
  The single transverse spin asymmetry data for the charged hadron production
 in pC 
and pCu interactions are presented. The measurements have been performed
 at  FODS-2 experimental setup using 40 GeV/c IHEP polarized proton beam.
 The hadron transverse momentum range is from 0.5 GeV/c up to 4 GeV/c.
 The data obtained off the nuclear targets are compared with the proton target
 data measured earlier with the same experimental setup and with the
data of other  experiments. 

\vskip 10mm

{\bf Key-words:} spin, asymmetry, hadron production,polarization, quarks, QCD.
\end{minipage}
\end{center}

\vskip 50mm
\begin{center}
{\it   Submitted to the XVII International Seminar on High Energy Physics \\
"Relativistic Nuclear Physics and Quantum Chromodynamics", \\
September 27 - October 2, 2004, Dubna, Russia}
\end{center}
%
\newpage
\section{Introduction}

 The experiments performed during the last 25 years show, that the single
 spin asymmetries
in hadron hadron interactions are much larger than expected from 
the naive pQCD.
  Also the hyperon transverse polarization is unexpectedly large in collisions
 of unpolarized hadrons. The  pertubative QCD predicts vanishing of single 
spin effects due to the vector nature of gluons and a small current quark mass.
There were no measurements so far of single spin asymmetries of charged
hadron production in the energy range between 22 and 200 GeV. 
We have measured the single-spin asymmetry $A_N$ of the inclusive charged
pion, kaon, proton and antiproton production cross sections at high
$x_T$ and high $x_F$ for a 40 $\gevp$ polarized proton beam incident 
on nuclei (C, Cu), where $A_N$ is defined as
 \begin{equation}
 A_{N} = { 1\over{P_{B}\cdot cos{\phi}} } \cdot
{ {N{\uparrow} - N{\downarrow}} \over {N{\uparrow} + N{\downarrow}} }, \quad
\end{equation}
where $P_B$ is the beam polarization, $\phi$ is the athimuthal angle of the 
production plane, ${N\!\!\uparrow}$ and ${N\!\!\downarrow}$ are event rates
for the beam spin up and down respectively. The measurements were
carried out at IHEP, Protvino in 2003. 
\section{Polarized beam and experimental setup}
The polarized protons are produced by the parity - nonconserving 
$\ensm{\Lambda}$ decays [1].
%
 The up or down beam transverse
polarization is achieved by the selection of decay protons with angles
near $90^{\circ}$ in the $\ensm{\Lambda}$ rest frame by
a movable collimator.  At the end of the beam line two  magnets 
 correct the vertical beam position on the spectrometer target for the two
beam polarizations. The intensity of the 40 $\gevp$ momentum
polarized beam on the spectrometer target  is $3\times 10^{7}$ ppp,
 $\Delta p/p= \pm 4.5$\%, the transverse polarization is $39^{+1}_{-3}$ \%,
 and the polarization direction
is changed each 18 min during 30 s. The beam intensity and the position
are measured by  ionization chambers and  scintillation
hodoscopes.
%
%

 Two Cherenkov counters 
 identify the beam particle
composition to control the background contamination.  At the spectrometer
magnet entrance there are two scintillation hodoscopes to measure
the vertical coordinates of the particles emitted from the target.
%
%
%
%
%
%

\begin{figure}[h]
\vskip -5mm
 \centerline{
 \includegraphics[width=90mm,height=70mm]{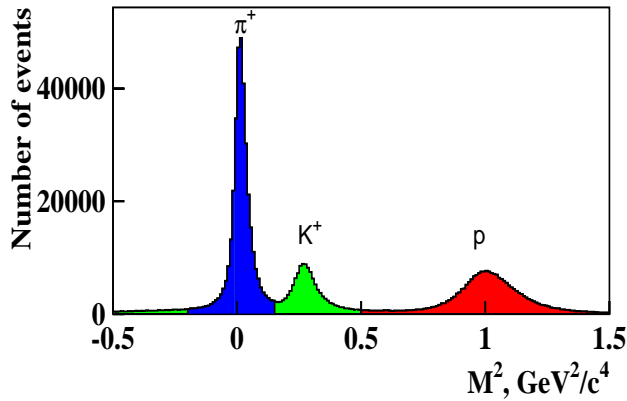}
}
\vskip -5mm
 \caption{Reconstructed hadron mass squared $M^{2}$ in the SCOCH spectrometer.
 \label{scochm}}
\end{figure}

  The measurements have been carried out with the FODS-2 [1] spectrometer. 
  It consists of an analyzing magnet,  drift chambers, the
  Cherenkov radiation spectrometer (SCOCH) for the particle
identification ($\pipm$ , $\PK^{\pm}$ , $\Pp$ and $\bar{\rm{p}}$), 
 the scintillation counters, and the
hadron calorimeters  to  trigger on the high energy hadrons. 
 Inside the magnet there is also a beam dump made of tungsten
and copper.  There are two arms which can be rotated around the target
center situated in front of the magnet to change the secondary
particle angle.
  The  Cherenkov radiation spectrometer
consists of a spherical mirror with the diameter 110 cm, 24
cylindrical lenses to focus the Cherenkov light on the hodoscope
photomultipliers. 
Measuring the particle velocity using the SCOCH and
its momentum in the magnetic field one can determine the particle square
mass $\ensm{\,{\rm M}}^2$ (Fig.~\ref{scochm}).  The SCOCHs are filled with
Freon~13 at 8~atm. 
%
%
%

 There are two threshold Cherenkov
counters using air at the atmospheric pressure inserted in the
magnet which are used for further improvement of particle identification.  
\section{Measurements}
In 1994 the study of the single spin asymmetry
($A_N$) in the inclusive charge hadron production was started using FODS-2:
\begin{equation}
\rm{p\!\uparrow + p(A) \rightarrow h^{\pm} + X},
\end{equation}
\begin{equation}
\rm{p\!\uparrow + p(A) \rightarrow h^{\pm}+ h^{\pm}  + X},
\end{equation}
 where $\rm{h^{\pm}}$ is a charged hadron
(pion, kaon, proton or antiproton).  The experimental program consists of 
 measuring the
charge hadron single spin asymmetry at high $x_T$ and $x_F$ in
$\Pp\Pp$ and $\Pp$A collisions to study the
asymmetry dependence on the quark flavors $\Pu$, $\Pd$, $\Ps$
 and kinematical variables.

The pilot measurements of $A_N$ for the charged hadrons carried out in
1994 with a hydrogen target for small $x_F$ [1]
 are presented below for the
comparison with the data obtained with  nuclei. The data for large
 $x_{F} \le 0.7$ were also measured in 2003. 

 The measurements of $A_{N}$ in
the range $-0.15 \le x_{F} \le 0.2$ and $0.5 \le p_{T} \le 4$ GeV/$c$ 
are carried out with symmetrical
arm positions at  angles of $\pm$ 160 mrad. The results for the two arms and
  the different values of magnetic field in the spectrometer are
averaged, which partially cancels systematical uncertaities connected with the
variation of the beam position in the vertical direction, the
intensity monitor and the apparatus  drift.
\section{Results}
\begin{figure}[h]
\vskip -15mm 
\centerline{\epsfxsize=7.4in\epsfbox{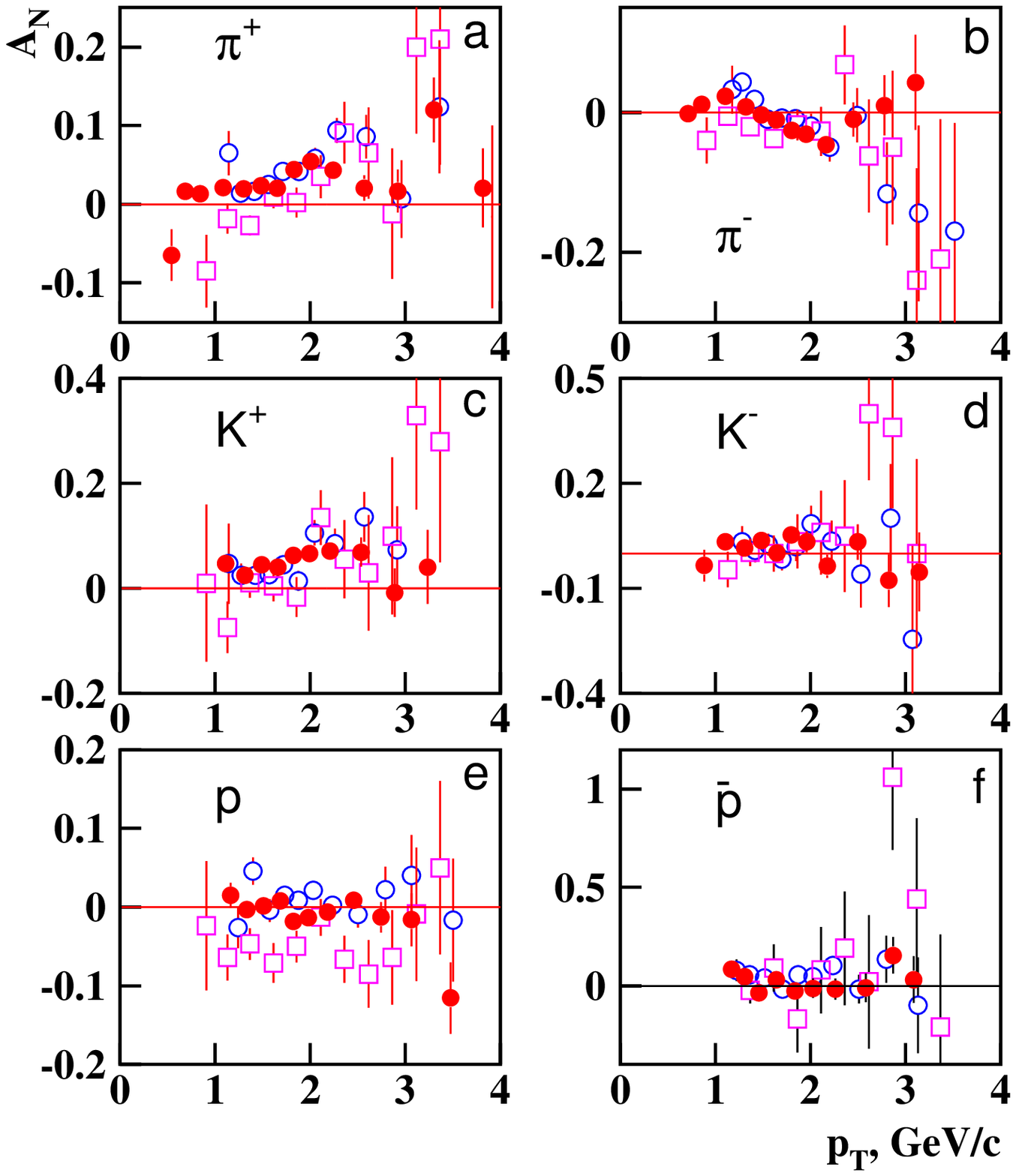}} 
\caption{$A_{N}$ dependence on  $p_T$ for 
$\rm{p\!\uparrow + p(A) \rightarrow h^{\pm} + X}$, where $\rm{h=\pip}$ (a),
 $\rm{\pim}$ (b),  $\rm{\PK^{+}}$ (c), $\rm{\PK^{-}}$ (d), $\Pp$ (e),
   $\rm{\bar{p}}$ (f). Closed cirles correspond to C target, open cirles - Cu,
 square - proton.   \label{asym}}
\end{figure}

The analyzing power results for six types of hadrons are shown
 in Fig.~\ref{asym}.  The errors quoted are statistical ones. 
This report presents the first data measured with two different nuclear 
targets. There were no data
 so far for charged hadrons with $p_{T} \ge $ 2.2 $\gevp$.

  In Fig.~\ref{asym}a  the $\pip$ 
meson production asymmetry is shown.  
Within the errors there is no difference
of $A_{N}$ for  both targets (C and Cu). $A_{N}$  for the nuclear targets in
the range $1 \le p_{T} \le 2$ $\gevp$ is approximately 4\% higher than
 for the hydrogen
target. For the central region such a difference can be connected with
the smaller portion of $\Pu$ quarks in the nuclear target containing
neutrons. 
Fragmentation of $\Pu$ quarks 
($\rm{u}\rightarrow \pip $)
from the polarized beam protons as well as  $\Pu$ quarks of the target
contribute to the asymmetry.
 Because the target protons are not polarized their
contribution in the central region reduces the measured polarization.
For nuclear targets containing less $\Pu$ quarks in comparison with $\Pd$
quarks the decrease of the asymmetry is not so substantial. Quark
scattering in nuclei must also lead to the decrease of the asymmetry due to 
the $p_{T}$ shift. 

 The asymmetry for $\pim$ meson production is presented 
in Fig.~\ref{asym}b. In the range $0.9 \le p_{T}
\le 1.6$ $\gevp$ it is about 4\% higher for the nuclear targets than for the
hydrogen target. For the central region such differences can be
connected to the larger proportion of $\Pd$ quarks in the nuclear targets.
 The major   contribution give $\Pd$ quarks ($\Pd \rightarrow \pim $ 
fragmentation) from the polarized beam protons and the target. For $\pim$
mesons in $\Pp\Pp$ collisions the asymmetry is therefore negative. 
Due to the large
contribution of the unpolarized target in the central regions the asymmetry
for nuclear targets is shifted into the positive region. 

 Fig.~\ref{asym}c shows
the asymmetry for $\PK^+$ production. There is no significant
difference in $A_N$ for the two
nuclear targets (C and Cu)  and $A_N$ is about
 3\% higher than for the hydrogen target. The reason of that can be 
the same as for  $\pip$ mesons.

In Fig.~\ref{asym}d  $A_N$ for $\PK^-$ mesons is presented.
 Within the errors there is no appreciable   difference in $A_{N}$ for
 all targets ($\Pp$, C and Cu) and $A_N$ is
 close to zero. This is expected because $\PK^-$ does not contain valence
 quarks from the polarized beam proton.

In Fig.~\ref{asym}e  the asymmetry for the  proton production is shown
 which is  close to zero for nuclear targets.
For the hydrogen target it is slightly negative.

 The asymmetry for antiproton production presented 
in Fig.~\ref{asym}f shows no
difference for all targets ($\Pp$, C and Cu) and is close to zero.
This result is expected  because the produced antiproton
does not contain valence quarks from the beam proton. Sea quarks in the
most models are expected to be unpolarized. The absolute values of analyzing
 power for antiprotons and $\PK^-$ mesons may be used as an estimate of
systematic bias ($\sim 4\%$) in the data in general.
\section{Discussion}
\begin{figure}[h]
\centering
\begin{tabular}{cc}
\begin{minipage}{75mm}
\includegraphics[width=70mm,height=80mm]{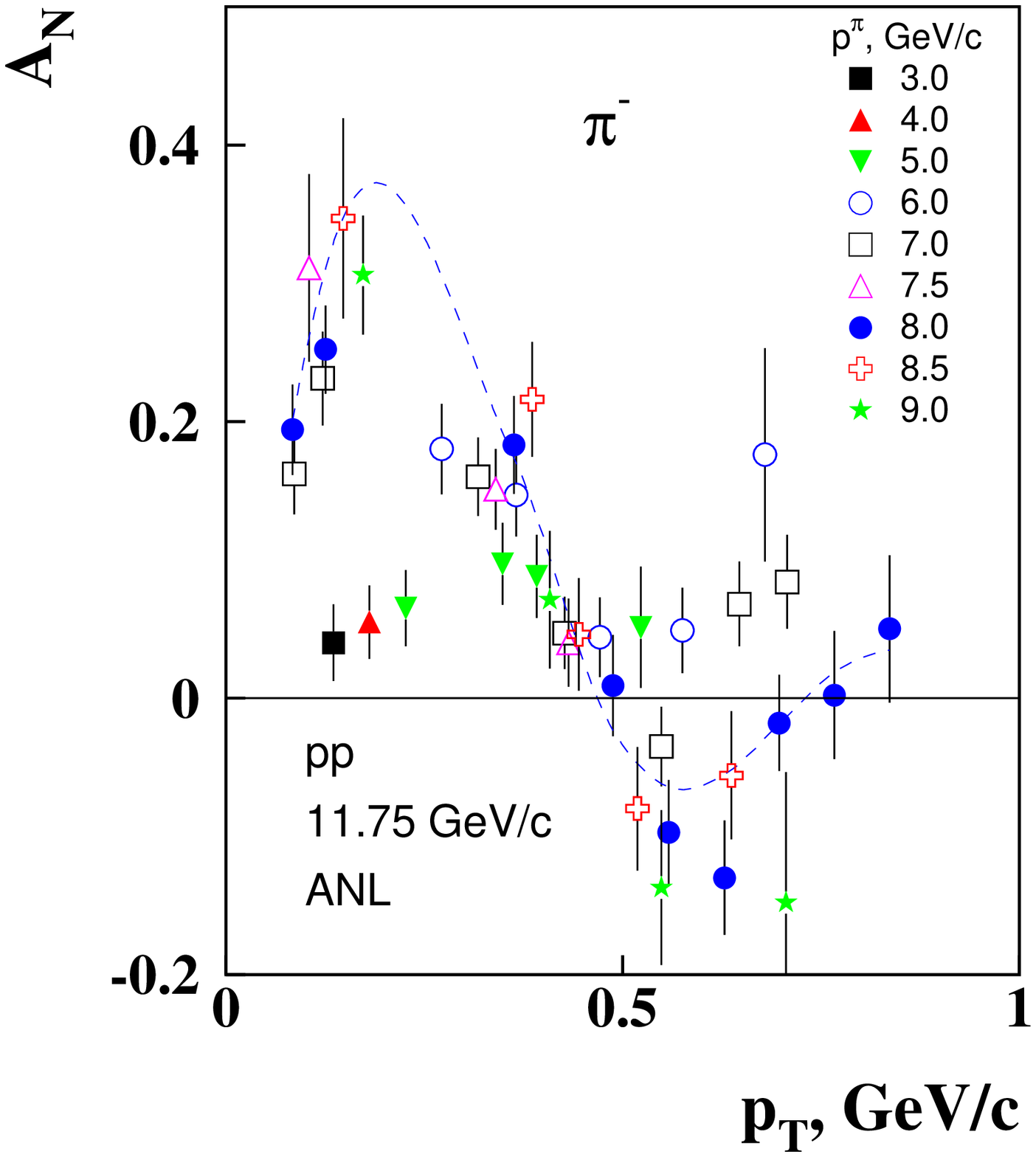}\caption{ $A_{N}$ dependence
 on  $p_T$ for $\rm{p\! \uparrow + p \rightarrow \pi^{-} + X}$ \cite{ANL}.
 Different symbols correspond to different secondary pion momenta. 
\label{ANL}} 
\end{minipage}
& 
\begin{minipage}{75mm}
\includegraphics[width=70mm,height=80mm]{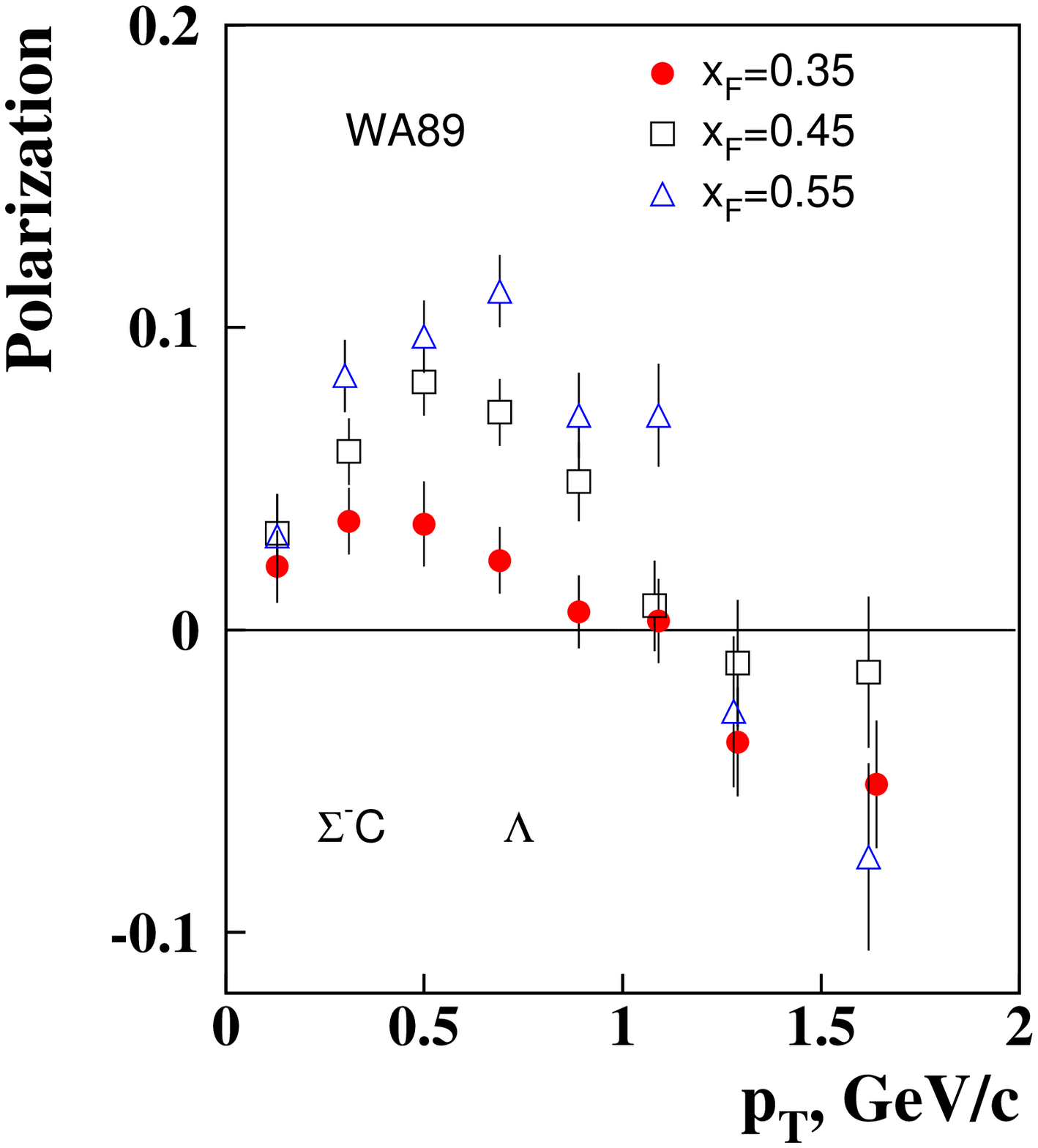}\caption{Polarization
 dependence on  $p_T$ for 
$\rm{\Sigma^{-} + C \rightarrow \Lambda + X}$ \cite{WA89}.
 Closed cirles correspond to $x_{F}=0.35$, 
open cirles - $x_{F}=0.45$,
 square - $x_{F}=0.55$.   \label{WA89}} 
\end{minipage} \\
\end{tabular}
\end{figure}

  It is instructive to compare 40 $\gevp$ data with the other data.
 There are very detailed ANL measurements performed at 11.75 GeV/c \cite{ANL}.
 There is a similarity in $p_{T}$ dependence of the analyzing power for
 negative pions at 40 $\gevp$ (Fig.~\ref{asym}b) 
and 11.75 $\gevp$ (Fig.~\ref{ANL}). 
In particular, the $A_{N}$ is rising
 with $p_{T}$, and then dropping back to zero or even negative values.
 The typical $p_{T}$ values of maximum 
and minimum positions are increasing with the beam energy.

 Another interesting similarity is between  $p_{T}$ dependence 
of the positive pion analyzing power (Fig.~\ref{asym}a) and 
the $\Lambda$ hyperon polarization in $\Sigma^{-}$ Carbon 
interactions \cite{WA89}. In Fig.~\ref{WA89} a change-over in $p_{T}$ 
dependence is seen. Such  behaviour has never been 
observed before in single spin effects for high energy inclusive reactions.
\section{Conclusion}
The analyzing powers were measured for $-0.15 \le x_{F} \le 0.2$ and
$0.5 \le p_{T} \le 4$  $\gevp$ in inclusive charged hadron production
off carbon and copper with 40  $\gevp$ polarized proton beam.
  Three features of the results can be stressed:

 (a) there is no significant  difference for the two nuclear targets (C, Cu);

 (b) for the positive charge mesons the asymmetry has a maximum at  $p_{T}=2.2$
$\gevp$ and decreases to zero at $p_{T}=2.9$ $\gevp$;

 (c) for hadrons not containing valence quarks from polarized protons
  ($\PK^-$ and $\rm{\bar{p}}$) the analyzing power is close to zero.

 The analysis of high $x_F$ data measured in 2003 is still under way.

 We are grateful  to the IHEP staff for their assistence with setting up
 the experiment,  and to the IHEP directorate for their support.

\end{document}